\documentclass[]{emulateapj}
\usepackage{epsfig}
\usepackage{multirow}

\defcitealias{2000AJ....120.3323B}{BK00}

\begin{document}
\title{Exploring the Outer Solar System with the ESSENCE Supernova Survey}
\author{
  A.~C.~Becker\altaffilmark{1},
  K.~Arraki\altaffilmark{1},
  N.~A.~Kaib\altaffilmark{1},
  W.~M.~Wood-Vasey\altaffilmark{2}, 
  C.~Aguilera\altaffilmark{3},
  J.~W.~Blackman\altaffilmark{4},
  S.~Blondin\altaffilmark{2},
  P.~Challis\altaffilmark{2},
  A.~Clocchiatti\altaffilmark{5},
  R.~Covarrubias\altaffilmark{6},
  G.~Damke\altaffilmark{3},
  T.~M.~Davis\altaffilmark{7,8},
  A.~V.~Filippenko\altaffilmark{9},
  R.~J.~Foley\altaffilmark{9},
  A.~Garg\altaffilmark{2,10},
  P.~M.~Garnavich\altaffilmark{11},
  M.~Hicken\altaffilmark{2,10},
  S.~Jha\altaffilmark{10,12},
  R.~P.~Kirshner\altaffilmark{2},
  K.~Krisciunas\altaffilmark{11,13},
  B.~Leibundgut\altaffilmark{14},
  W.~Li\altaffilmark{9},
  T.~Matheson\altaffilmark{15},
  A.~Miceli\altaffilmark{1},
  G.~Miknaitis\altaffilmark{16},
  G.~Narayan\altaffilmark{2,10},
  G.~Pignata\altaffilmark{17},
  J.~L.~Prieto\altaffilmark{18},
  A.~Rest\altaffilmark{3,10},
  A.~G.~Riess\altaffilmark{19,20},
  M.~E.~Salvo\altaffilmark{4},
  B.~P.~Schmidt\altaffilmark{4},
  R.~C.~Smith\altaffilmark{3},
  J.~Sollerman\altaffilmark{7, 21},
  J.~Spyromilio\altaffilmark{14},
  C.~W.~Stubbs\altaffilmark{2,10},
  N.~B.~Suntzeff\altaffilmark{3,13}, 
  J.~L.~Tonry\altaffilmark{22}, and
  A.~Zenteno\altaffilmark{3}
}

\altaffiltext{1}{Department of Astronomy, University of Washington, Box 351580, Seattle, WA 98195-1580.}
\altaffiltext{2}{Harvard-Smithsonian Center for Astrophysics, 60 Garden Street, Cambridge, MA 02138.}
\altaffiltext{3}{Cerro Tololo Inter-American Observatory, National Optical Astronomy Observatory (CTIO/NOAO), Colina El Pino, Casilla 603, La Serena, Chile.}
\altaffiltext{4}{The Research School of Astronomy and Astrophysics, The Australian National University, Mount Stromlo and Siding Spring Observatories, via Cotter Road, Weston Creek, PO 2611, Australia.}
\altaffiltext{5}{Pontificia Universidad Cat\'olica de Chile, Departamento de Astronom\'ia y Astrof\'isica, Casilla 306, Santiago 22, Chile.}
\altaffiltext{6}{Observatories of the Carnegie Institution of Washington, 813 Santa Barbara Street, Pasadena, CA 91101.}
\altaffiltext{7}{Dark Cosmology Centre, Niels Bohr Institute, University of Copenhagen, Juliane Maries Vej 30, DK-2100 Copenhagen \O, Denmark.}
\altaffiltext{8}{Department of Physics, University of Queensland, QLD, 4072, Australia}
\altaffiltext{9}{Department of Astronomy, University of California, Berkeley, CA 94720-3411.}
\altaffiltext{10}{Department of Physics, Harvard University, 17 Oxford Street, Cambridge, MA 02138.}
\altaffiltext{11}{Department of Physics, University of Notre Dame, 225 Nieuwland Science Hall, Notre Dame, IN 46556-5670.}
\altaffiltext{12}{Kavli Institute for Particle Astrophysics and Cosmology, Stanford Linear Accelerator Center, 2575 Sand Hill Road, MS 29, Menlo Park, CA 94025.}
\altaffiltext{13}{Department of Physics, Texas A\&M University, College Station, TX 77843-4242.}
\altaffiltext{14}{European Southern Observatory, Karl-Schwarzschild-Strasse 2, D-85748 Garching, Germany.}
\altaffiltext{15}{National Optical Astronomy Observatory, 950 North Cherry Avenue, Tucson, AZ 85719-4933.}
\altaffiltext{16}{Fermilab, P.O. Box 500, Batavia, IL 60510-0500.}
\altaffiltext{17}{Departamento de Astronomi\'a, Universidad de Chile, Casilla 36-D, Santiago, Chile}
\altaffiltext{18}{Department of Astronomy, Ohio State University, 4055 McPherson Laboratory, 140 West 18th Avenue, Columbus, OH 43210.}
\altaffiltext{19}{Space Telescope Science Institute, 3700 San Martin Drive, Baltimore, MD 21218.}
\altaffiltext{20}{Johns Hopkins University, 3400 North Charles Street, Baltimore, MD 21218.}
\altaffiltext{21}{Department of Astronomy, Stockholm University, AlbaNova, 10691 Stockholm, Sweden.}
\altaffiltext{22}{Institute for Astronomy, University of Hawaii, 2680 Woodlawn Drive, Honolulu, HI 96822.}

\begin{abstract} 

We report the discovery and orbit determination of 14 trans-Neptunian
objects (TNOs) from the ESSENCE Supernova Survey difference imaging
dataset.  Two additional objects discovered in a similar search of the
SDSS-II Supernova Survey database were recovered in this effort.
ESSENCE repeatedly observed fields far from the Solar System ecliptic
($-21^\circ < \beta < -5^\circ$), reaching limiting magnitudes per
observation of $I \approx 23.1$ and $R \approx 23.7$.  We examine
several of the newly detected objects in detail, including 2003
UC$_{414}$ which orbits entirely between Uranus and Neptune and lies
very close to a dynamical region that would make it stable for the
lifetime of the Solar System.  2003 SS$_{422}$ and 2007 TA$_{418}$
have high eccentricities and large perihelia, making them candidate
members of an outer class of trans-Neptunian objects.  We also report
a new member of the ``extended'' or ``detached'' scattered disk, 2004
VN$_{112}$, and verify the stability of its orbit using numerical
simulations.  This object would have been visible to ESSENCE for only
$\sim 2\%$ of its orbit, suggesting a vast number of similar objects
across the sky.  We emphasize that off-ecliptic surveys are optimal
for uncovering the diversity of such objects, which in turn will
constrain the history of gravitational influences that shaped our
early Solar System.


\end{abstract}
\keywords{surveys --- methods: data analysis --- Kuiper Belt} 

\section{Introduction}
\label{sec-intro}

The discovery of the accelerating universe in 1998 \citep[][; for a
review, see Filippenko 2005]{1998AJ....116.1009R, 1999ApJ...517..565P}
has given rise to a large number of next-generation surveys searching
for distant supernovae to probe the cosmological dark energy.  These
surveys are typically undertaken with wide-field imaging cameras to
ensure areal coverage broad enough to find significant numbers of
supernovae, and use moderate to large-aperture telescopes to probe
for faint supernovae at high redshifts.  A given supernova is
typically sampled every few days to resolve its brightness and color
evolution.

Within a given night, one of the most frequent contaminants to
supernova searches is foreground Solar System objects, which leave a
similar new-object signature in every image containing them.  In
addition, since supernova surveys tend to reach much deeper than
dedicated Solar System surveys, the majority of these moving objects
will be uncatalogued.  For this reason, multiple temporal observations
of a supernova candidate are required to verify its spatial
persistence before scheduling it for spectroscopic follow-up
observations.  Multiple images may be taken on a single night to
ensure that any Solar System objects show slight astrometric motion
(trans-Neptunian objects have reflex motions of $\sim$1$\arcsec$
hr$^{-1}$), or on different nights, allowing the Solar System object
to have moved significantly (instead proving to be a contaminant in
some other location).  These objects are typically ignored by the
surveys, but given the integrated amount of data available, provide
the opportunity for significant advances in our understanding of the
Solar System.

\section{Methods}
\label{sec-methods}

The observing strategy for the ESSENCE supernova survey is described
by \cite{2007ApJ...666..674M}.  These observations have been optimized
for the characterization of the dark energy equation-of-state parameter 
$w$ \citep[e.g.,][]{2003PhR...380..235P}.  In summary, the strategy was to
take two images of a given field per night using the Blanco 4-m
telescope plus MOSAIC-II imaging camera at the Cerro Tololo
Inter-American Observatory (CTIO).  One image was taken in the
$I$ band and the other in the $R$ band, typically separated by 
$\sim$60 min.  The exposure times lead to approximate limiting
$5\sigma$ magnitudes of $I \approx 23.1$ and $R \approx 23.7$.  The 
survey has thirty-two 0.36 deg$^2$ fields, each of which was observed
roughly every fourth night. 
ESSENCE images were obtained for 20~d around new moon for six years,
from 2002 to 2007, during 3 consecutive months, usually October
through December.  This observing cadence is serendipitously useful
for the study of TNOs.  It has sufficiently large intra-night spacing
to allow slight astrometric motion, yielding an instantaneous angular
velocity.  It also provides enough intra-month observations to recover
a given object several times per lunation, allowing us to link pairs
of observations that have consistent motion vectors.

ESSENCE uses a real-time difference imaging pipeline \cite[{\tt
Photpipe};][]{2002SPIE.4836..395S} that operates at the base camp of
CTIO.  Images are reduced and differenced immediately after
acquisition, and information on the detections found in the difference
images is posted to the internet for review by a team member.  Objects
clearly in motion are rejected from this visual analysis, and objects
not confirmed in follow-up observations are similarly ignored.  It is
this set of data that we wish to mine for distant Solar System
objects.

In this effort, we searched through {\it all} detections reported by
ESSENCE's {\tt Photpipe} difference imaging pipeline for the 6 seasons
of ESSENCE operations.  We kept all observations that were
positive-flux excursions, and which had a signal-to-noise ratio of at
least 5.  This yielded in total $3.7 \times 10^6$ independent
detections.  If we naively attempted to link all permutations of these
$N$ observations into tracks $M$ observations long, the problem would
scale as $N^M$.  This would very quickly become computationally
intractable.  It is primarily for this reason that such studies have
not been attempted in the past.  However, new methods of parsing and
organizing these data allow us to rapidly prune infeasible matches,
allowing computational scalings as fast as $N$\, log($N$)
\citep{2007Icar..189..151K}.

We used a prototype of the software developed by
\cite{2007Icar..189..151K} to link the pairs of $R$ and $I$-band
observations each night into $\sim 1$-hr ``tracklets,'' as well as
to link these tracklets across nights, into potential orbits called
``tracks.''  
%
%
For computational efficiency, we split the data by observing season
for the intra and inter-night linkages.  For intra-night linkages, we
required at least 2 detections whose separations implied angular
velocities less than $0.05^\circ$~d$^{-1}$, which would reject objects
at opposition and on circular orbits having semimajor axes $a$ $<$ 15
Astronomical Units (AU).  This process yielded $1.6 \times 10^5$
tracklets, which were next linked between nights.  For these
inter-night linkages, we allowed tracks with a maximum angular
velocity of $0.05^\circ$~d$^{-1}$, maximum angular acceleration of
$0.03^\circ$~d$^{-2}$, and supporting observations on at least 4
nights.  At time of maximum angular acceleration $90^\circ$ from
opposition, the acceleration cut would reject objects on circular
orbits with $a \lesssim 35$ AU.  However, the majority of our
observations were taken within $40^\circ$ of opposition, where this
cut would reject objects with $a \lesssim 20$ AU.  These particular
limits were chosen as a compromise between the goal of searching for
TNOs and the computational burden of fitting additional spurious
tracks.  This process yielded
$3.2 \times 10^6$ quadratic tracks as potential orbits.

We fit each track using the software of \citet[hereafter
\citetalias{2000AJ....120.3323B}]{2000AJ....120.3323B} to weed out
linkages that do not correspond to Keplerian motion.  We removed all
tracks with best-fit semimajor axes $a$ $<$ 10 AU, since the software
model uses a linear set of equations only valid for distant objects.
We rejected all fits whose $\chi^2$ per degree of freedom was greater
than 2.0.  Given each preliminary orbit, we searched again through the
difference imaging detections for matches on nights where there were
data in only one of the two passbands.  These additional points helped
to validate as well as extend each orbital arc.  This winnowing
process yielded 16 acceptable orbits with an average of 15
observations per object, and an average orbital arc of 50~d, excluding
%
%
6 objects that were detected in multiple seasons.  The
RMS deviations of our measured positions from the best-fit models is
approximately $0.1\arcsec$.

A summary of the objects detected and their orbital parameters is
given in Table~\ref{tab-params}.  We list the
\citetalias{2000AJ....120.3323B} fit parameters and uncertainties from
the ESSENCE data alone, including semi--major axis $a'$, eccentricity
$e'$, and inclination $i'$.  We include the $\chi^2$ per degree of
freedom of the fit and length of ESSENCE's orbital arc in years.  The
$\chi^2$ values are artificially small because the
\citetalias{2000AJ....120.3323B} software overestimates the
astrometric uncertainty per measurement at $0.2\arcsec$.  We also list
the most recent orbital parameters from the MPCORB database $a$, $e$,
and $i$, as well as the absolute magnitude $H$, defined as the
apparent visual magnitude at zero phase angle and 1 AU distance from
both the Earth and Sun.


\section{Results}
\label{sec-results}

While the yield from this search is modest in terms of the number of
objects detected, the search is noteworthy in that half of the ESSENCE
fields are significantly off the ecliptic ($-21^\circ < \beta <
-5^\circ$).  This provides a higher sensitivity to high-inclination
objects than normal ecliptic surveys.  As Table~\ref{tab-params}
shows, $\sim 70\%$ of our objects have inclinations greater than $10^\circ$.  
This is a larger fraction than that found in a similar search of
the SDSS-II Supernova Survey data ($\sim 40\%$) by \cite{SDSS-TNO},
and significantly larger than the fraction of high-inclination objects
in the known sample of all distant objects ($\sim 5\%$).


The ESSENCE observing strategy is significantly different than in
typical TNO surveys; its temporal cadence is designed to optimally
constrain lightcurves of distant supernovae as opposed to discover and
follow-up Solar System bodies \citep[e.g.][]{2006Icar..185..508J}.
The common wisdom borne of these past surveys is that at least two
oppositions worth of data are needed before one can compute a reliable
orbit or begin to distinguish between dynamical classes.  We
re-examine these presumptions to ascertain the reliability of our
single-opposition orbits.


The primary issue to be resolved is whether or not a single season of
data taken at ESSENCE's observing cadence is sufficient to distinguish
between different dynamical classes of objects.  
%
%
%
%
%
To examine the accuracy of our single-opposition orbits, we first
divide the data from our 6 multi-opposition objects 
into subsets delimited by observing season.  We then fit these subset
tracks with the \citetalias{2000AJ....120.3323B} software and compare
the subset fit parameters $a$, $e$, and $i$ to the solution from the
full fit, normalizing the difference by the associated uncertainty
from the subset fit.  We find that the software actually {\it
  overestimates} the uncertainties on single-opposition parameters,
which have a mean offset from their multi-opposition fits of $\sim 0.3
\sigma$.  By reducing the astrometric measurement uncertainties to a
more representative $0.1\arcsec$ we find mean offsets of $\sim 0.6
\sigma$.
%
%
The implication is that our single-opposition orbits are relatively
robust and that \citetalias{2000AJ....120.3323B} appear to do a
conservative job at assigning uncertainties to the orbital parameters.
%

%



The dynamical classification and interpretation of TNOs typically
requires numerical simulations of their nominal orbits, as well as the
orbits of an ensemble of clones that have orbits consistent with the
accumulated astrometry
\cite[e.g.,][]{2007Icar..189..213L,2008ssbn.book..275M}.  
Such an effort is beyond the scope of this paper.
%
%
However, qualitative classifications can be drawn from an object's
orbital parameters, with the caveat that some single-opposition orbits
may be significantly affected by assumptions inherent to the fitting
software and may change characteristics in a non-linear fashion with
additional observations.
Below we examine the dynamical implications of 2003 UC$_{414}$ (one
opposition), 2003 SS$_{422}$ (one opposition), 2007 TA$_{418}$ (two
oppositions), and 2004 VN$_{112}$ (two oppositions).

\subsection{2003 UC$_{414}$}
\label{sec-uc414}
The \citetalias{2000AJ....120.3323B} orbital parameters and those
extracted from the MPCORB database are in stark disagreement for 2003
UC$_{414}$, as seen from Table~\ref{tab-params}.  The source of this
discrepancy is unclear.  To resolve this issue, we turn to a third
independent package, {\tt OrbFit}, developed by
\cite{1999Icar..137..269M}.  Its orbital solution has $a = 25.9 \pm
0.1$ AU, $e = 0.08 \pm 0.02$, and $i = 26.4 \pm 0.4$ degrees, very
much in agreement with the \citetalias{2000AJ....120.3323B} solution,
which we adopt here.



The orbit of 2003 UC$_{414}$ is interesting because it has a low
eccentricity and is positioned nearly halfway between Uranus and
Neptune.  Given the strong gravitational perturbations caused by the
giant planets, this intuitively seems like a very unstable orbital
configuration.  In fact, there are only two known similar objects with
orbital arcs longer than two days : (160427) 2005 RL$_{43}$
\citep{SDSS-TNO} and 2000 CO$_{104}$.  Dynamical simulations suggest
that there are two islands of stability between Uranus and Neptune,
with $a \sim 24.6$ and 25.6 AU \citep{1997Natur.387..785H}.  The
dynamical lifetimes of objects in these regions is $\sim 10^9$ years.
Any confirmed members would provide additional constraints on models
of Solar System evolution that include violent dynamical instabilities
in the orbits of Uranus and Neptune
\citep[e.g.][]{2007arXiv0712.0553L}, which should depopulate these
regions.  Because of 2003 UC$_{414}$'s relatively short arc and
uncertain orbital parameters, more observations of this particular
object are necessary to ascertain if it lies within either of these
regions.

\subsection{2003 SS$_{422}$ and 2007 TA$_{418}$}
Both 2003 SS$_{422}$ and 2007 TA$_{418}$ have high-eccentricity (0.50
and 0.80, respectively), non-Neptune interacting ($q$ = 36.2 and 39.2
AU) orbits.  \cite{2003MNRAS.338..443E} have examined a similar set of
objects, selected by $a > 49.9$ AU and $q > 30.9$ AU, integrating
their orbits and those of clones for 4.5 Gyr.  They find that a
substantial portion of such high-eccentricity objects do not reach the
near-Neptune region in the age of the Solar System, making the
scattered-disk population an unlikely origin for these objects.  There
appears to be a soft cutoff of $q \approx 35$ AU between stable and
unstable behavior.  Both 2007 TA$_{418}$ and 2003 SS$_{422}$ are near
this threshold, and must be analyzed in a similar manner to determine
their stability.  2003 SS$_{422}$ is particularly interesting in this
regard, having a larger semimajor axis and eccentricity than any
object in the \cite{2003MNRAS.338..443E} study other than 2000
CR$_{105}$ \citep{2002Icar..157..269G}.

\subsection{2004 VN$_{112}$}

2004 VN$_{112}$ is one of our better-constrained objects, with an
orbital arc of 420~d.  Its high inclination ($25.6^\circ$)
indicates that it would preferably have been detected by surveys
observing far off the ecliptic, where the object is found when near
perihelion.
The large semimajor axis (315 AU) and eccentricity (0.85) provide a
perihelion $q$ of 47.2 AU, a circumstance that places it beyond the
dynamical control of any major body currently known in our Solar
System.  2004 VN$_{112}$ likely represents a new member of the
``extended'' scattered disk \cite[ESD; e.g.,][]{2002Icar..157..269G}.
ESD objects have perihelia that detach them from dynamical
interactions with Neptune, typically defined as $q > 40$ AU
\citep{2007Icar..189..213L}.

To ascertain its orbital stability, we generated 1000 clones of 2004
VN$_{112}$ from a multivariate normal distribution incorporating the
covariances between orbital parameters derived from the
\cite{1999Icar..137..269M} software.  We integrated these for 1 Gyr
using the modified version of the SWIFT-RMVS3 integrator
\citep{1994Icar..108...18L} as outlined in \cite{2007arXiv0707.4515K}.
In these integrations, we include the gravitational effects of the
Sun, the four giant planets, passing field stars, as well as the Milky
Way tide.  After 1 Gyr of evolution, we find that the orbits of our
clones are relatively unchanged.  To be strongly altered by the
perturbations from Neptune, the perihelion of 2004 VN$_{112}$ would
have to migrate inside $\sim 40$ AU, and in our simulations we find
$\left<\left(\Delta q\right)^{2}\right>^{1/2} = 1.7$ AU for our clones
after $10^{9}$ yrs with no bias toward inward or outward migration.
Alternatively, this orbit could also be significantly modified by
Galactic tides if its semimajor axis grows beyond $\sim 1000$ AU.
This does not occur for any of our clones, with $a = 392$ AU being the
largest semimajor axis attained at the end of our simulation.
%
%
Given these results, we can conclude that this orbit is stable for the
history of the Solar System.

The perihelion of 2004 VN$_{112}$ is very near the 2:1 orbital
resonance with Neptune.  An intriguing possibility is that it was
placed on its (currently stable) orbit by a primordial member of the
Solar System that was subsequently ejected due to resonant
interactions with Neptune.  As detailed in simulations by
\cite{2006ApJ...643L.135G}, this rogue planet scenario tends to
produce higher-inclination objects at smaller semimajor axis.
Comparing 2004 VN$_{112}$ to the ensemble of detached TNOs defined by
\cite{2007Icar..189..213L}, we find that 2004 VN$_{112}$ has the
second-largest semimajor axis after (90377) Sedna, suggesting it
should have an inclination between $12^\circ$ and $23^\circ$.  Its
inclination of nearly $25.6^\circ$ (with a fitted uncertainty of
$0.004^\circ$) is inconsistent with a monotonic decrease in
inclination with increasing semimajor axis for the ESD.  However,
there will be some variance around the relationship, making this a
non-definitive constraint.
%
An alternative scenario is that the ESD was formed through
perturbations by passing stars, which yields increasing inclinations,
eccentricities, and perihelia at larger semimajor axis
\cite[e.g.,][]{2004AJ....128.2564M}.



While it is possible that 2004 VN$_{112}$ was a ``lucky'' find, we
proceed with an estimate of the ESD extent with the caveat that this
object may not faithfully represent the entire population.
2004 VN$_{112}$ was detected 0.8 mag from the limit of the ESSENCE
survey, and 0.3 AU from perihelion.  We estimate that such an object
would be visible for only $2\%$ of its orbit.  Given ESSENCE's areal
coverage, a rough estimate of the total number of similar objects or
brighter across the entire sky is $\sim 10^5$.  The exact number is a
function of the unknown inclination distribution for these objects.
Simulations of the scattered disk by \cite{2004MNRAS.355..935M}
suggest that the majority ($70-90\%$) of objects are found at
inclinations lower than $25^\circ$.  However, the current inclination
distribution of the ESD is unknown.  For our order--of--magnitude
estimates here, we adopted a cutoff at $40^\circ$.  If we further
assume an albedo of 0.05 (yielding a diameter of 300 km given its
absolute $H$-band magnitude of 6.4), and a power-law cumulative size
distribution with an index of 3, this implies a total number of
objects on similar (i.e., detached) orbits, and greater than 100~km in
size, of $10^{6-7}$.  This is similar to the estimates of
\cite{2002Icar..157..269G} based upon their detection of 2000
CR$_{105}$.

\section{Conclusions}
\label{sec-conclusions}

We report on a data-mining effort that resulted in the discovery and
orbital determination of 14 new trans-Neptunian bodies by the ESSENCE
Supernova Survey.  Only two previously known objects were seen, a high
ratio of discovery that highlights the utility and novelty of the
search.  Each object was detected multiple times over the span of
approximately 3 months, with several objects recovered in multiple
seasons of the survey.  All objects had sufficient data to receive
provisional designations from the Minor Planet Center.

Our sensitivity to high-inclination objects was higher than most
surveys due to our repeated visits to off-ecliptic fields.  We found a
substantial number of objects with both large inclinations and high
eccentricities.  These bodies could only have received such orbits
through interactions with a scattering body.  2004 VN$_{112}$ stands
out in this regard, having an orbit that detaches it from
gravitational interactions with the major bodies of our current Solar
System.  We have verified that this orbit is stable on 1 Gyr
timescales by numerically integrating $10^3$ clones.  As a member of
the extended scattered disk, 2004 VN$_{112}$ provides an additional
constraint on theories of external perturbations and early evolution
that shaped today's Solar System.  In particular, its orbital
parameters appear inconsistent with a model in which currently
detached objects were previously scattered by a rogue planet.
Revealing the overall trend of inclination with semimajor axis will
help resolve the origin of the ESD, a study that suggests more
observations at even higher ecliptic latitudes.  Our detection of 2004
VN$_{112}$ suggests that there are $10^{6-7}$ objects greater than
100~km in size in the ESD, a vast number whose ensemble properties
will help us understand the early evolution of our Solar System.

The success of this study demonstrates that vast amounts of
astronomical survey data may be usefully and efficiently mined for
Solar System objects.  This is a direct result of advances in the
fields of image subtraction \citep{1998ApJ...503..325A},
data-reduction pipelines \citep{2002SPIE.4836..395S}, and data-linking
techniques \citep{2007Icar..189..151K}.  The recent suggestion
\citep{2007RPPh...70..883W} that dark energy studies are bad for
astronomy provides a helpful warning not to let those programs become
focused exclusively on a single goal.  Our work shows that a deep
survey carried out to constrain the dark energy equation of state also
contains a wealth of information that can be successfully mined for
other valuable science.  
%
%
%
%
Observations well outside the ecliptic plane will detect a variety of
objects that can provide clues to the evolution of the Solar System,
making high ecliptic latitude a region ripe for discovery.

\acknowledgements We thank L. Jones and A. Puckett for useful
discussions, and J. Kubica for assistance in using the object-linking
software.  This publication makes use of the MPCORB database provided
by the Minor Planet Center.
Based in part on observations obtained at the Cerro Tololo
Inter-American Observatory, which is operated by the Association of
Universities for Research in Astronomy, Inc. (AURA) under cooperative
agreement with the National Science Foundation (NSF).  The survey is
supported by the NSF through grants AST-0443378, AST-057475,
AST-0606772, and AST-0607485.  A.C. acknowledges grant FONDECYT
1051061 from CONICYT, Chile.  A.R. thanks the NOAO Goldberg Fellowship
Program for its support.  G.P acknowledges support by the Proyecto
FONDECYT 3070034.


\begin{thebibliography}{22}
\expandafter\ifx\csname natexlab\endcsname\relax\def\natexlab#1{#1}\fi

\bibitem[{{Alard} \& {Lupton}(1998)}]{1998ApJ...503..325A}
{Alard}, C. \& {Lupton}, R.~H. 1998, \apj, 503, 325

\bibitem[{{Becker} {et~al.}(2008)}]{SDSS-TNO}
{Becker}, A.~C., {et~al.} 2008, in preparation

\bibitem[{{Bernstein} \& {Khushalani}(2000)}]{2000AJ....120.3323B}
{Bernstein}, G. \& {Khushalani}, B. 2000, \aj, 120, 3323

\bibitem[{{Emel'yanenko} {et~al.}(2003){Emel'yanenko}, {Asher}, \&
  {Bailey}}]{2003MNRAS.338..443E}
{Emel'yanenko}, V.~V., {Asher}, D.~J., \& {Bailey}, M.~E. 2003, \mnras, 338,
  443

\bibitem[{{Gladman} \& {Chan}(2006)}]{2006ApJ...643L.135G}
{Gladman}, B. \& {Chan}, C. 2006, \apjl, 643, L135

\bibitem[{{Gladman} {et~al.}(2002){Gladman}, {Holman}, {Grav}, {Kavelaars},
  {Nicholson}, {Aksnes}, \& {Petit}}]{2002Icar..157..269G}
{Gladman}, B., {et~al.} 2002, Icarus, 157, 269

\bibitem[{{Holman}(1997)}]{1997Natur.387..785H}
{Holman}, M.~J. 1997, \nat, 387, 785

\bibitem[{{Jones} {et~al.}(2006)}]{2006Icar..185..508J}
{Jones}, R.~L. {et~al.} 2006, Icarus, 185, 508

\bibitem[{{Kaib} \& {Quinn}(2007)}]{2007arXiv0707.4515K}
{Kaib}, N.~A. \& {Quinn}, T. 2007, ArXiv e-prints, 707

\bibitem[{{Kubica} {et~al.}(2007){Kubica}, {Denneau}, {Grav}, {Heasley},
  {Jedicke}, {Masiero}, {Milani}, {Moore}, {Tholen}, \&
  {Wainscoat}}]{2007Icar..189..151K}
{Kubica}, J., {et~al.} 2007, Icarus, 189, 151

\bibitem[{{Levison} \& {Duncan}(1994)}]{1994Icar..108...18L}
{Levison}, H.~F. \& {Duncan}, M.~J. 1994, Icarus, 108, 18

\bibitem[{{Levison} {et~al.}(2007){Levison}, {Morbidelli}, {Van Laerhoven},
  {Gomes}, \& {Tsiganis}}]{2007arXiv0712.0553L}
{Levison}, H.~F., {et~al.} 2007, ArXiv e-prints, 712

\bibitem[{{Lykawka} \& {Mukai}(2007)}]{2007Icar..189..213L}
{Lykawka}, P.~S. \& {Mukai}, T. 2007, Icarus, 189, 213

\bibitem[{{Miknaitis} {et~al.}(2007){Miknaitis}, {Pignata}, {Rest},
  {Wood-Vasey}, {Blondin}, {Challis}, {Smith}, {Stubbs}, {Suntzeff}, {Foley},
  {Matheson}, {Tonry}, {Aguilera}, {Blackman}, {Becker}, {Clocchiatti},
  {Covarrubias}, {Davis}, {Filippenko}, {Garg}, {Garnavich}, {Hicken}, {Jha},
  {Krisciunas}, {Kirshner}, {Leibundgut}, {Li}, {Miceli}, {Narayan}, {Prieto},
  {Riess}, {Salvo}, {Schmidt}, {Sollerman}, {Spyromilio}, \&
  {Zenteno}}]{2007ApJ...666..674M}
{Miknaitis}, G., {et~al.} 2007, \apj, 666, 674

\bibitem[{{Milani}(1999)}]{1999Icar..137..269M}
{Milani}, A. 1999, Icarus, 137, 269

\bibitem[{{Morbidelli} {et~al.}(2004){Morbidelli}, {Emel'yanenko}, \&
  {Levison}}]{2004MNRAS.355..935M}
{Morbidelli}, A., {Emel'yanenko}, V.~V., \& {Levison}, H.~F. 2004, \mnras, 355,
  935

\bibitem[{{Morbidelli} \& {Levison}(2004)}]{2004AJ....128.2564M}
{Morbidelli}, A. \& {Levison}, H.~F. 2004, \aj, 128, 2564

\bibitem[{{Morbidelli} {et~al.}(2008){Morbidelli}, {Levison}, \&
  {Gomes}}]{2008ssbn.book..275M}
{Morbidelli}, A., {Levison}, H.~F., \& {Gomes}, R. 2008, {The Dynamical
  Structure of the Kuiper Belt and Its Primordial Origin} (The Solar System
  Beyond Neptune), 275--292

\bibitem[{{Padmanabhan}(2003)}]{2003PhR...380..235P}
{Padmanabhan}, T. 2003, \physrep, 380, 235

\bibitem[{{Perlmutter} {et~al.}(1999){Perlmutter}, {Aldering}, {Goldhaber},
  {Knop}, {Nugent}, {Castro}, {Deustua}, {Fabbro}, {Goobar}, {Groom}, {Hook},
  {Kim}, {Kim}, {Lee}, {Nunes}, {Pain}, {Pennypacker}, {Quimby}, {Lidman},
  {Ellis}, {Irwin}, {McMahon}, {Ruiz-Lapuente}, {Walton}, {Schaefer}, {Boyle},
  {Filippenko}, {Matheson}, {Fruchter}, {Panagia}, {Newberg}, {Couch}, \& {The
  Supernova Cosmology Project}}]{1999ApJ...517..565P}
{Perlmutter}, S., {et~al.} 1999, \apj, 517, 565

\bibitem[{{Riess} {et~al.}(1998){Riess}, {Filippenko}, {Challis},
  {Clocchiatti}, {Diercks}, {Garnavich}, {Gilliland}, {Hogan}, {Jha},
  {Kirshner}, {Leibundgut}, {Phillips}, {Reiss}, {Schmidt}, {Schommer},
  {Smith}, {Spyromilio}, {Stubbs}, {Suntzeff}, \&
  {Tonry}}]{1998AJ....116.1009R}
{Riess}, A.~G., {et~al.} 1998, \aj, 116, 1009

\bibitem[{{Smith} {et~al.}(2002){Smith}, {Rest}, {Hiriart}, {Becker}, {Stubbs},
  {Valdes}, \& {Suntzeff}}]{2002SPIE.4836..395S}
{Smith}, C., {et~al.} 2002, in Survey and Other Telescope
  Technologies and Discoveries. Edited by Tyson, J. Anthony; Wolff, Sidney.
  Proceedings of the SPIE, Volume 4836, pp. 395-405 (2002)

\bibitem[{{White}(2007)}]{2007RPPh...70..883W}
{White}, S.~D.~M. 2007, Reports of Progress in Physics, 70, 883

\end{thebibliography}

\begin{table}[thbp]
  \setlength{\abovecaptionskip}{0pt}
  \setlength{\belowcaptionskip}{10pt}
  \begin{center}
    \caption{Summary of Orbital Parameters for the ESSENCE Sample}
    \label{tab-params}
          {\footnotesize
            \begin{tabular}{l|ccccc|cccc}
              \hline\hline
              {\em Object} & {\em a' (AU)}  & {\em e'}  & {\em i' (deg)}  & {\em $\chi^2$/d.o.f.} & {\em dT (yr)} & {\em a}  & {\em e}  & {\em i}  & {\em H} \\
              \hline

              2003 UC$_{414}$\footnotemark[1]     &  26.0 (0.1)   &  0.09 (0.06) &  26.4 (0.1)   &  0.06 &  0.16 &
                                     44.9       &  0.64        &  25.9           & 8.3 \\ 

              2006 TK$_{121}$     &  38.5 (0.7)   &  0.21 (0.04) &  27.27 (0.02)   &  0.10 &  0.25 &
                                     38.4         &  0.21        &  27.30          & 8.1 \\

              2003 WN$_{193}$     &  39.4 (0.4)   &  0.253 (0.007) &  21.62 (0.01)   &  0.07 &  0.11 &
                                     39.4         &  0.253         &  21.63        & 8.5 \\

              2003 SR$_{422}$     &  40.11 (0.04) &  0.056 (0.005) &  23.914 (0.002) &  0.10 &  1.30 &
                                     40.07        &  0.055         &  23.939         & 7.1 \\

              2007 TZ$_{417}$     &  41.6 (0.1)   &  0.14 (0.01)   &  22.280 (0.004) &  0.25 &  1.14 &
                                     41.6       &  0.14        &  22.310         & 7.5 \\

              2005 SE$_{278}$\footnotemark[2] &  42.31 (0.02)   &  0.110 (0.002) &  6.892 (0.001) &  0.07 &  1.24 &
                                     42.34        &  0.111         &  6.894         & 7.1 \\

              2006 QQ$_{180}$\footnotemark[2] &  42.7 (9.3) &  0.21 (0.36) &  9.4 (0.2)     &  0.12 &  0.09 &
                                     42.3       &  0.18        &  9.4          & 6.8 \\  

              2007 VJ$_{302}$     &  43.1 (0.2)   &  0.065 (0.002) &  8.70 (0.01)   &  0.07 &  1.20 &
                                     43.1       &  0.066         &  8.73          & 6.8 \\

              2003 WO$_{193}$     &  44.2 (16.6)  &  0.38 (0.40) &  6.626 (0.003)  &  0.09 &  0.08 &
                                     38.6         &  0.19        &  6.628          & 8.3 \\

              2007 VK$_{302}$     &  46.7 (5.7) &  0.11 (0.69) &  26.3 (0.7)     &  0.15 &  0.09 &
                                     43.5       &  0.08        &  28.1           & 7.0 \\

              2007 TD$_{418}$     &  52.8 (7.0) &  0.33 (0.16) &  15.091 (0.001) &  0.15 &  0.11 &
                                     45.2       &  0.13        &  15.095         & 7.9 \\

              2007 TC$_{418}$     &  53.6 (8.3) &  0.34 (0.24) &  10.6 (0.2)     &  0.13 &  0.11 &
                                     43.1       &  0.11        &  11.3           & 7.6 \\

              2007 TA$_{418}$     &  72.8 (1.6) &  0.51 (0.01)   &  21.962 (0.001) &  0.11 &  1.24 &
                                     72.7       &  0.50        &  21.964         & 7.2 \\

              2007 TB$_{418}$     &  90.0 (56.9) &  0.67 (0.25) &  6.55 (0.02)   &  0.36 &  0.16 &
                                     55.3        &  0.39        &  6.57          & 5.8 \\

              2003 SS$_{422}$     &  203 (46) &  0.81 (0.05) &  16.78 (0.04)   &  0.16 &  0.21 &
                                     196      &  0.80        &  16.81          & 7.1 \\

              2004 VN$_{112}$     &  319 (6) &  0.852 (0.003) &  25.550 (0.004) &  0.04 &  1.15 &
                                     319     &  0.852         &  25.580         & 6.4 \\

              \hline\hline
            \end{tabular}
          }
  \end{center}
  
  Note -- Orbital parameters for the ESSENCE TNO sample.  We include
  initial orbital parameters and uncertainties derived using the
  \citetalias{2000AJ....120.3323B} software : semimajor axis $a'$, shown in
  AU; orbital eccentricity $e'$; and orbital inclination $i'$ in
  degrees.  We include the $\chi^2$ per degree--of--freedom from the
  fit, as well as the orbital arc length in years.  We next list the
  current orbital parameters taken from the MPCORB database provided
  by the Minor Planet Center, including the absolute magnitude, $H$,
  defined as the apparent visual magnitude at zero phase angle and 1
  AU distance from both the Earth and Sun.  \\

  {\small
  1 -- As outlined in Section~\ref{sec-uc414}, the
  \citetalias{2000AJ....120.3323B} fit is preferred for object 2003
  UC$_{414}$.

  2 -- 2005 SE$_{278}$ and 2006 QQ$_{180}$ were previously discovered
  by the SDSS-II Supernova Survey \citep{SDSS-TNO}.

  }

\end{table}

\end{document}